\documentclass[useASM]{mn2e}

\usepackage{graphicx}

\begin{document}

\title{The ``amplitude'' parameter of Gamma-Ray Bursts and its
implications for GRB classification }

\author[L\"{u} et al.]
{Hou-Jun L\"{u}$^{1}$ , Bing Zhang$^{1}$ \thanks{E-mail:
lhj@physics.unlv.edu, zhang@physics.unlv.edu }, En-Wei
Liang$^{2,1}$, Bin-Bin Zhang$^{3}$ and Takanori Sakamoto$^{4}$
\\
 $^1$Department of Physics and Astronomy, University of Nevada, Las Vegas, NV 89154, USA\\
 $^2$Department of Physics and GXU-NAOC Center for Astrophysics and Space Sciences, Guangxi University, Nanning 530004, China\\
 $^3$Center for Space Plasma and Aeronomic Research (CSPAR), University of Alabama in Huntsville, Huntsville, AL 35899, USA\\
 $^4$Department of Physics and Mathematics, Aoyama Gakuin University, Sagamihara-shi Kanagawa 252-5258, Japan\\}
 \maketitle

\label{firstpage}
\begin{abstract}
Traditionally gamma-ray bursts (GRBs) are classified in
the $T_{90}$-hardness ratio two-dimensional plane into
long/soft and short/hard GRBs. In this paper, we suggest to add
the ``amplitude'' of GRB prompt emission as the third dimension
as a complementary criterion to classify GRBs, especially those
of short durations. We define three new parameters $f$, $f_{\rm
eff}$ and $f_{\rm eff,z}$ as ratios between the
measured/simulated peak flux of a GRB/pseudo-GRB and the flux
background, and discuss the applications of these parameters to
GRB classification. We systematically derive these parameters
to find that most short GRBs are likely not ``tip-of-iceberg''
of long GRBs. However, one needs to be cautious if a short GRB
has a relatively small $f$ (e.g. $f<1.5$), since the chance for
an intrinsically long GRB to appear as a ``disguised'' short
GRB is higher. Based on avaialble data, we quantify the
probability of a disguised short GRB below a certain $f$ value
is as $P (<f)\sim 0.78^{+0.71}_{-0.4} f^{-4.33\pm 1.84}$. By
progressively ``moving'' a long GRB to higher redshifts through
simulations, we also find that most long GRBs would show up as
rest-frame short GRBs above a certain redshift.
\end{abstract}

\section {Introduction}
Traditionally, gamma-ray bursts (GRBs) are classified based on
duration ($T_{90}$) and hardness ratio (HR) of their prompt
gamma-ray emission. In the CGRO/BATSE era, GRBs were classified
into two categories in the $T_{90}$-HR two-dimensional plane
(Kouveliotou et al. 1993) with a rough separation in the
duration dimension at $T_{90} \sim 2$ s. Long GRBs are
typically soft while short GRBs are typically hard, so that the
two classes cluster in two regions in the $T_{90}$-HR plane.
Such a distribution is energy-dependent and
instrument-dependent (e.g. Qin et al. 2013; Zhang et al. 2012).
A third, intermediate class has been suggested by various
authors based on the duration criterion alone (e.g. Mukherjee
et al. 1998; Horvath 1998; Hakkila et al. 2000; Horvath et al.
2010).

Broad-band afterglow observations of long GRBs reveal that
their host galaxies are typically irregular galaxies with
intense star formation (Fruchter et al. 2006). Some long GRBs
are firmly associated with Type Ib/c supernova (e.g. Hjorth et
al. 2003; Stanek et al. 2003; Campana et al. 2006; Pian et al.
2006; Xu et al. 2013). This strongly suggests that they are likely related to
the deaths of massive stars, and the ``collapsar'' model has
been widely accepted to be the standard paradigm for long GRBs
(Woosley 1993; MacFadyen \& Woosley 1999). Detections of
afterglows and host galaxies of short GRBs in the Swift
(Gehrels et al. 2004) era have advanced our understanding of
their physical origin. Some short GRBs are found to be
associated with nearby early-type galaxies with little star
formation (Gehrels et al. 2005; Bloom et al. 2006; Barthelmy et
al. 2005; Berger et al. 2005), or have a large offset from the
host even if they are associated with star forming galaxies
(e.g. Fox et al. 2005; Fong et al. 2010). Deep upper limits of
their supernova signals are obtained (Kann et al. 2011, Berger 2014
and references therein). This points towards an origin that does
not involve a massive star. The leading scenario is mergers of
two neutron stars (Pacz\'ynski 1986; Eichler et al. 1989) or
mergers of a neutron star and a black hole (Pacz\'ynski 1991b).
There is no evidence that the intermediate third class forms a
physically distinct population of GRBs.

Further observations revealed a more complicated picture,
suggesting that duration is no longer a reliable indicator of
the physical origin of a GRB. The detections of two nearby
long-duration GRBs without association of a supernova, i.e. GRB
060614 ($T_{90}\sim100 $ s at $z=0.125$) and GRB 060505
($T_{90}=4$ s at $z=0.089$), cast doubts on that all long GRBs
are of a massive star origin (Gehrels et al. 2006; Gal-Yam et
al. 2006; Fynbo et al. 2006; Della Valle et al. 2006). On the
other hand, some properties of GRB 060614 (e.g. short spectral
lag, Gehrels et al. 2006) and the large offset from the star
forming region in the host (Gal-Yam et al. 2006) are consistent
with being a compact star origin. Zhang et al. (2007b) showed
that if GRB 060614 were somewhat less energetic, it would
appear as quite similar to GRB 050724, which is the ``smoking
gun'' short GRB (with extended emission) that suggests a
compact star origin (Barthelmy et al. 2005; Berger et al.
2005). Later, several high-$z$ GRBs with the rest frame
duration $T_{90}/(1+z)$ shorter than 2 s were discovered: GRB
080913 at $z=6.7$ with $T_{90}=8$ s (Greiner et al. 2009), GRB
090423 at $z=8.2$ with $T_{90}=10.3$ s (Tanvir et al. 2009;
Salvaterra et al. 2009), and GRB 090429B at $z=9.4$ with
$T_{90} = 5.5$ s (Cucchiara et al. 2011), but various arguments
suggest that they are of a massive star origin (Zhang et al.
2009). Later, more traditional short GRBs are found to be
likely of a massive star origin. For example, GRB 090426, at
$z=2.609$, is found to have an observed BAT band duration
$T_{90} = 1.2\pm 0.3$s and a rest frame duration
$T_{90}/(1+z)\sim 0.33$ s, but its other properties are fully
consistent with being of a massive star origin (Levesque et al.
2010; Xin et al. 2011; Th\"one et al. 2011).

In view of these complications, Zhang (2006) and Zhang et al.
(2007b) suggested to classify GRBs physically into Type II
(massive star origin) and Type I (compact star origin). Zhang
et al. (2009) studied the statistical properties of the Type II
and Type I Gold Samples, and found that although the Type II
Gold sample tracks the bulk of long GRBs well, the Type I Gold
sample is not a good representative of the short GRBs. They
suggested a set of multi-wavelength criteria to diagnose the
physical origin of GRBs (see also Kann et al. 2011), and
suspected that some, maybe most high-redshift high-luminosity
short GRBs would be of a Type II origin. This conclusion
was later also drawn by several groups independently based on
very different arguments (Virgili et al. 2011; Cui et al. 2012;
Bromberg et al. 2012).

Even though the multi-wavelength criteria can give more
definite clues about the origin of a GRB, they are not
available promptly after the trigger of the GRB. Some criteria
that carry most weight (e.g. supernova signature, host galaxy
information) need late, deep optical observations. It is still
useful to apply the prompt gamma-ray data to dig out more
information, which may be helpful to infer the physical origin
of a GRB. For example, in L\"u et al. (2010), we have proposed
a new observational parameter $\varepsilon$ defined by
$E_{\gamma,iso}/E_{p,z}^{5/3}$, where $E_{\gamma,iso}$ is the
burst isotropic gamma-ray energy and $E_{p,z}$ is the
rest-frame spectral peak energy. This parameter has a cleaner
bimodal distribution, and the two types of burst classified
with the $\varepsilon$ criterion match the physical
classification scheme (Type I vs. Type II) better. This method
still needs the redshift information.

In this paper, we propose to add a third dimension
``amplitude'' into consideration to classify GRBs using the
prompt gamma-ray data (see a preliminary discussion in Zhang
2012). The motivation is to study the possibility that a real
long GRB may be observed as a ``short'' one if the majority of
emission episode is too faint to be detected above the
background. We call this the ``tip-of-iceberg''
effect\footnote{In the early BATSE era, some authors had
introduced the effective amplitude parameters such as
$V/V_{max}$ or $C_{max}/C_{min}$ to perform statistical
analyses, but the purpose of their studies was to test for the
uniformity of the GRB spatial distribution (e.g. Schmidt et al.
1988; Paczynski 1991a).}. To quantify this effect, in Sect. 2,
we introduce a new ``amplitude parameter'' $f$, and study the
distribution of {\em Swift} GRBs in the three-dimensional
($T_{90} -{\rm HR}-f$) space. In Sect. 3, we introduce an
effective amplitude parameter $f_{\rm eff}$ to discuss the
range of amplitude if a long GRB is observed as ``short'' due
to the tip-of-iceberg effect. We compare the range of $f$
distribution of short GRBs and the $f_{\rm eff}$ distribution
of long GRBs and suggest a confusion regime of $f$ where an
observed short GRB may be in fact long. In Sect.4, we define a
parameter $f_{\rm eff,z}$ by ``moving'' GRBs with known
redshift to higher redshifts through simulations until they
become ``rest-frame short'' GRBs. We take GRB 080319B as an
example, and show that long GRBs can become rest-frame short
GRBs at high enough redshifts, but with a moderately large $f$.
We show that this is consistent with the three highest-$z$
GRBs: 080913, 090423 and 090429B. We draw conclusions in Sect.
5 with some discussion.

\section{The amplitude parameter $f$}

In the previous $T_{90}$-HR two-dimensional diagram, the
amplitude information of GRBs is missing. Some GRBs can be very
bright, while some others can be faint and barely above the
threshold. A bright burst can have more emission episodes
emerging above the background, so for a same observed $T_{90}$, a
fainter burst may be intrinsically longer than a brighter
burst. So this third dimension, i.e. the ``amplitude'', carries
important information and should be introduced in GRB
classification studies. Such a fluence truncation effect has
been studied extensively in the past (e.g. Koshut et al. 1996;
Bonnell et al. 1997; Hakkila et al. 2000; Schmidt 2001).

Here we quantify such an effect by defining an amplitude
parameter
\begin{equation}
f \equiv \frac{F_{p}}{F_{B}},
\label{f}
\end{equation}
where $F_{p}$ is the 1-second peak flux on the gamma-ray
emission lightcurve, and $F_{B}$ is the average background flux
of the burst. Both fluxes are in units of count rate.

We systematically process the {\em Swift} Burst Alert Telescope
(BAT) GRB data to extract lightcurves.  We developed an IDL
script to automatically download and maintain all the {\em
Swift} BAT data. We use the standard HEAsoft tools (version
6.12) to process the data. By running {\em bateconvert} from
the HEASoft software release,  we obtain the energy scale for
the BAT events. The lightcurves are extracted by running {\em
batbinevt} (Sakamoto et al. 2007). For each burst, we calculate
the cumulative distribution of the source count using the
arrival times of a fraction of 5\% and 95\% of the total counts
to define $T_{90}$ (see Fig 5). The time bin size is fixed to
64 ms for all the bursts. Background is extracted using
two time intervals, one before and one after the burst. By
fitting the background as a Poisson noise, one can obtain its
standard deviation. The error of $f$ is derived from the error
of $F_B$ based on error propagation.

Our sample includes the GRBs detected by {\em Swift} BAT from
December 2004 to December 2011. We only selected 437 GRBs with
S/N ratio higher than 5, which include 395 long GRBs and 42
short GRBs. Among them, 182 have redshift measurements. For
each GRB, we fit the background flux level $F_{B}$ using the
time intervals before and after the burst. This background is
burst-dependent, but is around a value of 8000 cts/s. For a
small fraction (6.8\%) of the bursts, the background before and
after the burst is uneven. This is because some bright hard
X-ray sources could be entering or exiting the BAT field of
view during the slew. For these cases, we fit the background
before and after the burst with a straight line with a slope.
$F_{B}$ is defined by fitted background flux at the peak
time\footnote{This flux level is usually slightly higher than
the ``true'' background level due to the source contamination.
However, this is not a concern for our analysis, since we are
investigating the tip-of-iceberg effect with respect to the
background at the detection time.}. Figure 1 shows the
histogram of $F_{B}$ for all the GRBs in our sample.

The $f$ values of the GRBs in our sample are presented at
http://grb.physics.unlv.edu/f/data.txt. The $T_{90}-{\rm HR}-f$
3-dimensional distribution diagram of {\em Swift} GRBs is shown
in Fig.2. Long and short GRBs are denoted as black and white
symbols. The projections in the $T_{90}-{\rm HR}$, $T_{90}-f$
and ${\rm HR}-f$ planes are denoted in red, green, and blue
colors, respectively, with long and short GRBs denoted by the
filled and open symbols, respectively. In Fig.3a, we show 1D
distribution ($T_{90}$ and $f$), and 2D ($T_{90}-f$) diagram
with different symbols denoting different types of GRBs: gray
for long GRBs, red for short GRBs, blue for short GRBs with
extended emission ($T_{90}$ calculated by excluding the
extended emission), purple for the three ``rest-frame short''
($T_{90}/(1+z) < 2$ s) high-$z$ GRBs, black for other
``rest-frame short'' GRBs, and two special GRBs, 090426 and
060614, are marked separately.

The distributions of the $f$-parameter for both long and short
GRBs are presented in Fig.4a. As expected, Most bursts are
clustered around small $f$ values, and only a small fraction of
bursts have $f>3$. The $f$ distribution can be roughly fit as a
power law function, i.e. $N(f) \propto f^{-a}$, with $a \sim
3.54$ for long GRBs and $a \sim 1.66$ for short GRBs. The mean
value of $f$ is $\bar f=1.48$ for long GRBs and $\bar f=1.82$
for short GRBs. The largest $f$ values for both long and short
GRBs are around 10. The relative paucity of small $f$ for short
GRBs may be understood as a selection effect (Sakamoto et al.
2008, 2011): Short GRBs are detected via ``rate triggers'',
which require a relatively large $f$ value to meet the trigger
criterion. On the other hand, long GRBs can be caught via
``imaging triggers'' near the threshold, so that they can be
detected with lower $f$ values close to unity.

Although the average value $f$ of long GRBs is smaller than
short GRBs, and the $N(f) \propto f^{-a}$ slope of the two
populations are considerably different, one cannot significantly
improve the duration classification scheme with the introduction
of the $f$ value. As shown below, when introducing the next
parameter $f_{\rm eff}$, one can gain useful information to judge
the true duration category of a GRB, especially for short GRBs.

\section{Effective amplitude $f_{\rm eff}$ of long GRBs, and
 short-GRB confusion}

A long GRB may be confused as a short GRB if only its brightest
spikes with duration shorter than 2 s are above the background.
To quantify such a tip-of-iceberg effect, we define an
``effective amplitude'' of a {\em long} GRB as
\begin{equation}
f_{\rm eff} \equiv \frac{F^{\prime}_{p}}{F_{B}}.
\label{feff}
\end{equation}
Here $F^{\prime}_p$ is the 1-second peak flux of a pseudo
GRB, which is re-scaled down for multiplying by a factor
$\epsilon$($\epsilon<1$) from an original GRB lightcurve until
its signal above the background has a duration $T_{\rm 90,eff}$
just shorter than 2 s. The physical meaning of the pseudo GRB
is an otherwise identical GRB at the same redshift, except that
the amplitude is lower by a factor $\epsilon$. Since a short
GRB has $T_{90}$ shorter than 2 s, if one defines a $f_{\rm
eff}$ parameter for a short GRB, it is identical to $f$. So we
only define $f_{\rm eff}$ for long GRBs.

Technically, the $f_{\rm eff}$ parameter of a long GRB is
measured based on the following procedure. 1. We extract the
lightcurve of an observed GRB following the standard procedure
with a time bin 64 ms; 2. We ``re-scale'' down the observed
lightcurve to reduce the flux at each time bin by multiplying
the flux by a factor $\epsilon$ ($\epsilon<1$) for each time bin,
and make a ``signal'' of a pseudo-GRB. 3. We simulate a Poisson
background based on the extracted background information (the
mean flux and standard deviation), and add this background to
the ``signal'' and derive an ``observed'' lightcurve of the
pseudo GRB; 4. For this simulated ``observed'' lightcurve, we
apply the standard ``curve of growth'' method by accumulating
net fluence above the back ground (e.g. von Kienlin et al. 2014).
The duration $T_{90}$ of the pseudo-GRB is obtained through
measuring the time interval between $5\%$ and $95\%$ fluence;
5. We progressively multiply by a factor
$\epsilon_{i}$($\epsilon_{i}<1$) with the original light curve,
each time record $T_{90}$ until the derived $T_{90}$ of the
pseudo GRB is below 2 s. Record the $f$ value of this pseudo
GRB and define it as $f_{\rm eff}$.

Figure 5 shows the long GRB 050525A as an example. The original
burst is shown in Fig.5a, which has an $f =9.43$. Figure 5b
shows a pseudo GRB after re-scaling it down by a factor of
$\epsilon=0.06$. The signal (thin black curve in Fig.5b) is
below the background level $F_{B}$ (the gray curve). The sum of
the signal and background gives a new ``observed'' lightcurve
(the orange curve) of the pseudo GRB, whose $T_{90}$ is
measured through the curve of growth method. Only the main peak
is within the $5\%-95\%$ window. The measured $T_{90}$ is just
shorter than 2 s. We then measure the $f$ value of this pseudo
burst, which is the effective amplitude of the original burst.
For this example, one measures $f_{\rm eff} =1.53$.

Figure 3b gives the 1D distributions of $f_{\rm eff}$, and the
$T_{90} - f_{\rm eff}$ distribution of long GRBs together with the
$T_{90} - f$ distribution of short GRBs in our sample. The
$f_{\rm eff}$ values of long GRBs are systematically smaller
than the $f$ values of short GRBs. The $f_{\rm eff}$
distribution histogram of long GRBs is also shown in Fig.4a,
which has a mean value $\bar f_{\rm eff} =1.24$, and the
steepest slope $a =8.04\pm 1.23$ as compared with $f$
distributions of long and short (see inset of Fig.4a).

One immediate conclusion from Fig.3b and Fig 4a is that
the distribution of $f_{\rm eff}$ of long GRBs is very
different from the $f$ distribution of short GRBs. Most short
GRBs have larger $f$ values than the $f_{\rm eff}$ values of
long GRBs. This suggests that the majority of short GRBs are
{\em not} tip-of-iceberg of long GRBs. Instead, they reflect
the intrinsically short duration of the central engine.
Nonetheless, at smaller $f$ values for short GRBs, confusion
would appear since some long GRBs may show up as ``disguised''
short GRBs due to the tip-of-iceberg effect. In Fig.4b, we
present the cumulative probability distribution of $f$ for
short GRBs and $f_{\rm eff}$ for long GRBs. It is clearly shown
that most long GRBs have small $f_{\rm eff}$ values, e.g. $\sim
95\%$ below 1.5. In contrast, only $\sim 30\%$ short GRBs have
$f<1.5$.

In order to quantify the chance probability of disguised
short GRBs, we carry out a Monte Carlo simulation. Since the
observed short GRBs may include both intrinsic and disguised
short GRBs, we assume an $f$ distribution $N(f) \propto f^{-\alpha}$
for the intrinsic short GRBs, with the slope $\alpha$ taken as
a parameter to be constrained by the data. We then simulate $10^4$
short GRBs whose $f$ distribution follows this distribution.
Next, we simulate a certain amount of disguised short GRBs
whose $f$-distribution satisfies the $f_{\rm eff}$
distribution of long GRBs. The observed short GRBs should
be a superposition of the intinsic and disguised short GRBs.
In order
to calibrate the two population, we notice that there are 7
observed short GRBs that have $f < 1.5$, and one of them
(GRB 090426) is a disguised short GRB (Levesque et al. 2010;
Xin et al. 2011; Th\"one et al. 2011) with $f=1.48$. The chance
probability for a disguised short GRB at $f \leq 1.5$ is therefore
$P(f<1.5) \sim 1/7 \sim 0.142$. With this calibration, we
obtain the ``observed'' short GRB sample by superposing the
simulated intrinsic and disguised short GRB samples.
We require that
$f$ distribution of this ``observed sample'' satisfies the
observed $f$ distribution, whose slope is $\sim 1.66$.
We find that the $\alpha$ value of the simulated intrinsic
short GRBs is only slightly shallower, with $\alpha \sim 1.61$.
This is understandable, since essentially all the observed short
GRBs at $f>1.5$ are intrinsic ones, and they define the slope
of the $f$-distribution of the intrinsic short GRB sample.
After reaching consistency with the data, we track the
fraction of intrinsic and disguised short GRBs in the
total simulated sample to
map the chance probability of disguised short GRB below
any $f$ value. This probability function reads
\begin{equation}
 P (<f)\sim 0.78^{+0.71}_{-0.4} f^{-4.33\pm 1.84}
\label{P}
\end{equation}
Since the $f$ and $f_{\rm eff}$ distribution indices have
errors, the chance probability in Eq.(\ref{P}) also have
errors. The coefficient error and the index error are
correlated. All the relations in any case
allow $P(f<1.5) =0.142$ (see
Fig.4c). One can see that the chance probability for
contamination can reach 78\% near $f=1$. So for detected short
GRBs with a small $f$ value (say $f<1.5$), one should be
cautious to draw conclusion about the duration category of the
GRB.

It is interesting to note that GRB 060614 (Gehrels et al.
2006), the peculiar long GRB without supernova association, has
$f_{\rm eff} = 1.75$. This means that its tip-of-iceberg still
has a large $f$ to be consistent with the short GRB $f$
distribution. Indeed, by scaling it down, it looks like a short
GRB with extended emission (Zhang et al. 2007b). Our analysis
again supports the Type I (compact star) origin of this GRB.

\section{The $f_{\rm eff,z}$ parameter and ``rest-frame short'' GRBs}

Some long GRBs have a rest-frame duration $T_{90}/(1+z) < 2$ s.
The three GRBs with the highest redshifts, i.e. GRB 080913
(Greiner et al. 2009), GRB 090423 (Tanvir et al. 2009;
Salvaterra et al. 2009), and GRB 090429B (Cucchiara et al.
2011) are all of this type, but likely have a Type II (massive
star) origin based on the multi-wavelength criteria (Zhang et
al. 2009). It would be very interesting to investigate whether
this is also due to the tip-of-iceberg effect.

In order to check such a possibility, we define a third
parameter
\begin{equation}
f_{\rm eff,z} \equiv \frac{F^{\prime}_{p,z}}{F_{B}}.
\label{feffz}
\end{equation}
Here $F'_{p,z}$ is the 1-second peak flux of a pseudo GRB,
which is generated by ``moving'' the original GRB to progressively
higher redshifts until the rest-frame duration $T_{90}/(1+z)$
becomes shorter than 2 s. A GRB, when moved to a higher
redshift, would usually have a shorter rest frame duration,
although the observed duration may not shrink due to time
dilation (Kocevski \& Petrosian 2013). In principle, it would
always reach the ``rest-frame-short'' phase before completely
disappearing beneath the background. It would be interesting to
investigate the critical redshift $z_c$ above which a burst
appears as rest-frame-short.

Technically, moving a GRB with known redshift to higher
redshifts is not straightforward. One needs to reduce the
time-resolved spectra of the GRB, derive the correct spectral
parameters, and perform a proper $k$-correction to the spectrum
in order to obtain the BAT-band light curve of the pseudo GRB.

To carry out such an exercise, for each GRB with redshift
measurement, we first apply {\em Xspec} to conduct a
time-dependent spectral analysis to the raw data. We dissect
the lightcurve into multiple time bins, with the bin size
self-adjusted to allow a signal-to-noise ratio S/N$>5$, so that
the spectral parameters can be constrained. A typical GRB
spectrum, if the observational band is wide enough, can be
described as the Band function (Band et al. 1993; Abdo et al.
2009; Zhang et al. 2011). In order to perform a proper $k$
correction, ideally one should know the Band spectral
parameters $\alpha$, $\beta$ and $E_p$. However, since the BAT
band is narrow, for most GRBs the spectra can be only fit by a
cutoff power law or a single power law (Sakamoto et al. 2008,
2011). We therefore apply the following procedure to estimate
the Band spectral parameters: 1. If a burst was also detected
by {\em Fermi} GBM or {\em Konus} Wind, we adopt the spectral
parameters measured by those instruments. 2. For those bursts
that were not detected by other instruments but can be fit with
a cutoff power law, we adopt the derived $\alpha$ and $E_p$
parameters, and assume a typical value $\beta = -2.3$. 3. For
those GRBs that could only be fit with a single power law, we
have to a derive $E_p$ using an empirical correlation between
the BAT-band photon index $\Gamma^{\rm BAT}$ and $E_p$, as
derived previously for Swift GRBs (Sakamoto et al. 2009; Zhang
et al. 2007a,b; Virgili et al. 2012). The typical parameters
$\alpha = -1$, $\beta=-2.3$ are adopted to perform the
simulations.

We note that moving a GRB to a higher $z$ is
effectively observing the rest-frame spectra in a higher energy
band given the same observed BAT band. The spectral parameters
$\beta$ and $E_p$ are therefore essential. These parameters are
unfortunately usually not available for {\em Swift} GRBs.
So our pseudo GRBs should be considered only as simulated GRBs
rather than the original GRBs being moved to higher redshifts.
In any case, such a simulation can serve the purpose of
investigating the tip-of-iceberg selection effect. A similar
simulation was carried out by Kocevski \& Petrosian (2011).

Given the spectral parameters $\alpha$, $\beta$ and $E_p$ of a
particular GRB with known redshift $z$, we use the following
procedure to simulate the pseudo GRB. First, we calculate the
time-dependent bolometric burst luminosity using
\begin{equation}
L(t)=4\pi D^{2}_{L}(z) F(t) k,
\end{equation}
where $F(t)$ is the BAT-band, time-dependent flux, $D_{L}(z)$
is the luminosity distance to the source at the redshift $z$,
and the $k$-correction factor corrects the BAT-band ($15-150$
keV) flux to a wide band in the burst rest frame ($1-10^4$ keV
in this analysis), i.e.
\begin{equation}
k=\frac{\int^{10^4/{1+z}}_{1/{1+z}}EN(E)dE}{\int^{150}_{15}EN(E)dE}.
\end{equation}
Here $N(E)$ is the time-dependent Band photon spectrum. To
calculate $D_{L}(z)$, the concordance cosmology parameters
($H_0 = 71$ km s$^{-1}$ Mpc$^{-1}$, $\Omega_M=0.30$, and
$\Omega_{\Lambda}=0.70$) are adopted.

Next, we apply the spectral model to calculate the BAT-band
flux for a pseudo GRB at redshift $z'$. We keep the bolometric
luminosity as a constant, and derive the BAT band flux using
\begin{equation}
F'(t') = \frac{L(t)} {4\pi D^{2}_{L}(z^{\prime}) k^{\prime}},
\end{equation}
where
\begin{equation}
t^{\prime}=\frac{1+z^{\prime}}{1+z} t,
\end{equation}
$D_{L}(z^{\prime})$ is the luminosity distance to the source at
redshift $z^{\prime}$, and
\begin{equation}
k^{\prime}=\frac{\int^{10^4/{1+z^{\prime}}}_{1/{1+z^{\prime}}}E' N(E')dE'}
{\int^{150}_{15}E' N(E')dE'}.
\end{equation}
Here $N(E')$ is the observed photon number spectrum of the
pseudo GRB. The spectrum is still a Band function with the same
$\alpha$ and $\beta$ values. The only difference is that the
peak energy is now shifted to
$E^{\prime}_{p}=E_{p}(1+z)/(1+z^{\prime})$. We then add the
background $F_B$ and its fluctuation based on simulation,
and re-calculate $T_{90}$ of the pseudo GRB for each $z'$
following the same procedure to derive $f_{\rm eff}$. We then
calculate the rest-frame duration using $T_{90}/(1+z)$.

By progressingly increasing $z'$, we
identify a critical redshift $z_c$ beyond which $T_{90}/(1+z) <
2$ s is satisfied. The peak flux of the pseudo GRB at $z_c$ is
used to define $f_{\rm eff,z}$. We continue to increase
the redshift, until the entire GRB disappears below the
background. We record this redshift as $z_{\rm max}$. The
redshift range $(z_c, z_{\rm max})$ is then where a rest-frame
short GRB is observed.

The parameter $f_{\rm eff,z}$ depends on several parameters,
such as $F(t)$ (which further depends on spectral parameters
$\alpha$, $E_p$, $\beta$ or $\Gamma^{\rm BAT}$),
and $F_B$. We have introduced the error of each measurable, and
properly derive the error of $f_{\rm eff,z}$ through error
propagation.

As an example, we take the ``naked-eye'' GRB 080319B (Racusin
et al. 2008) as the original burst and perform the simulation.
The results are shown in Fig.6. The time-integrated
$\gamma$-ray spectrum is well fit using a Band function, with
$E_{p}=675\pm 22$ keV. The time-resolved of spectra are well
obtained, with $E_{p}$ evolving from $\sim$740 keV to $\sim$540
keV. The rest-frame isotropic energy release is $E_{\rm iso} =
(1.14\pm0.09)\times10^{54}$ erg in the source frame $1-10^4$
keV band (Racusin et al. 2008, Amati et al. 2008).

We apply the above method to simulate pseudo GRBs with
increasing redshifts. The lightcurves of the pseudo GRBs are
presented in Fig.6a. Different colors denote different
redshifts. From top to bottom, the redshifts are: $z=0.937$
(original), 1, 2.3, 2.8, 3.6, 4.5, 5.1, and the critical
redshift is $z_c=5.53$. As shown in Fig.6a, the peak flux of
the pseudo GRBs become progressively lower as $z$ increases,
and the observed durations initially become longer (due to time
dilation) but later shrink (due to tip-of-iceberg effect). The
rest-frame duration $T_{90}/(1+z)$ is found to decrease with
redshift, similar to track with a smooth broken power-law
(Fig.6b). At $z=z_c = 5.53$, $T_{90}/(1+z)$ becomes shorter
than 2 s. We derive $f_{\rm eff,z} =1.41$. The burst is no
longer detectable at $z=z_{\rm max}=5.92$.

We carry out the same exercise for all the {\em Swift} GRBs
with known redshifts. The $T_{90}/(1+z)-f_{\rm eff,z}$ diagram
is presented in Fig.3c. We can see that $f_{\rm eff.z}$ are all
below $\sim$1.7. It is interesting to note that the three
highest-$z$ GRBs (080913, 090423 and 090429B) and other
rest-frame-short GRBs all have $f$ values within this range.
This suggests that they are simply the tip-of-iceberg of long
GRBs. This conclusion is consistent with their Type II origin
as derived from multi-wavelength arguments (Zhang et al. 2009).
In Fig.7, we plot the histograms of $z_c$ and $z_{\rm max}$ of
all the GRBs in our analysis, and compare them with the $z$
distribution of the detected rest-frame short GRBs. It is found
that they are generally consistent with each other. The
discrepancy in the high-$z$ end (the distribution does not
fully include the highest $z$ GRB) may be due to the
uncertainty of the high-energy spectra used in our simulations.

If the rest-frame short GRBs are the tip-of-iceberg of long
GRBs, then the extended emission episodes (``icebergs''
themselves) may show up in the softer X-ray band. To test this
possibility, in Fig.8 we simulate the expected XRT band
lightcurve of a pseudo naked-eye GRB 080319B at $z=z_c=5.53$
(black). The same $k$-correction method has been applied. This
is compared against the XRT-band lightcurves of the three
highest-$z$ GRBs (green for GRB 080913, blue for GRB 090423,
and red for GRB 090429B), as well as the original XRT-band data
of GRB 080319B (gray). It is seen that the XRT lightcurve of
the pseudo GRB has an extended flaring episode extending to
$\sim 200$ s followed by a steep decay, which is similar to the
case of GRB 090423.

\section{Conclusions and discussion}

In this paper, we propose to add ``amplitude'' as the third
dimension as a complementary criterion to study GRB classification
using the prompt emission
data. We introduced three parameters, $f$ (Eq.(\ref{f})),
$f_{\rm eff}$ (Eq.(\ref{feff})), and $f_{\rm eff,z}$
(Eq.(\ref{feffz})), to describe the amplitude of the original
GRB and some simulated pseudo GRBs. We find the following
interesting results:
\begin{itemize}
 \item The $f$ parameters for both long and short GRBs are
     distributed between 1 and about 10 as a rough power
     law. The paucity of low-$f$ short GRBs may be
     understood as a trigger selection effect.
 \item The $f$ parameter of many short GRBs is larger than
     the $f_{\rm eff}$ parameter of long GRBs. This
     suggests that most short GRBs are likely intrinsically
     short, and {\em not} simply the tip-of-iceberg of long
     GRBs.
 \item  There is a confusion regime as $f$ is small
     (e.g. $<1.5$) for short GRBs, since intrinsically long
     GRBs may show up as disguised short GRBs due to the
     tip-of-iceberg effect. GRB 090426 is such an example.
     Through simulations, we derive the chance probability
     of disguised short GRBs as a function of $f$ for short
     GRBs below a certain $f$ value (Eq.[\ref{P}]). The
     contamination becomes significant below $f\sim 1.5$,
     and can reach as large as $\sim 78\%$ at $f \sim 1$.
     This raises caution to judge the duration category of
     a short GRB with $f<1.5$.
 \item When long GRBs are moved to high redshifts, they are
     likely observed as rest-frame short GRBs due to the
     ``tip-of-iceberg'' effect. These rest-frame short GRBs
     are supposed to have a low amplitude $f < 1.7$. The
     observed three highest-$z$ GRBs and other rest-frame
     short GRBs all have such a low amplitude. So they are
     consistent with being tip-of-iceberg of long GRBs.
\end{itemize}

\section{Acknowledgements}
We thank the anonymous referee for helpful comments, and
acknowledge the use of the public data from the {\em Swift}
data archive. This work was supported by NASA NNX10AD48G,
NNX14AF85G, and NSF AST-0908362 (BZ). EWL acknowledges support
from the National Natural Science Foundation of China under
grants No. 11063001, 10873002 and 11025313, the National Basic
Research Program (``973" Program) of China (Grant
2009CB824800), Special Foundation for Distinguished Expert
Program of Guangxi, the Guangxi SHI-BAI-QIAN project (Grant
2007201). BBZ acknowledges support by NASA SAO G01-12102X.


\begin{figure*}
\includegraphics[angle=0,scale=1]{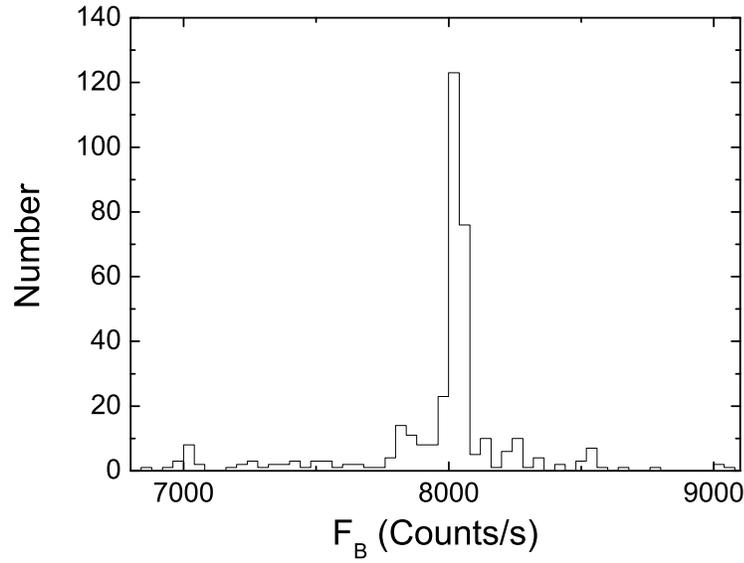}
\hfill\center \caption{The the histogram of $F_{B}$ for all the
GRBs in our sample.}
\end{figure*}


\begin{figure*}
\includegraphics[angle=0,scale=1]{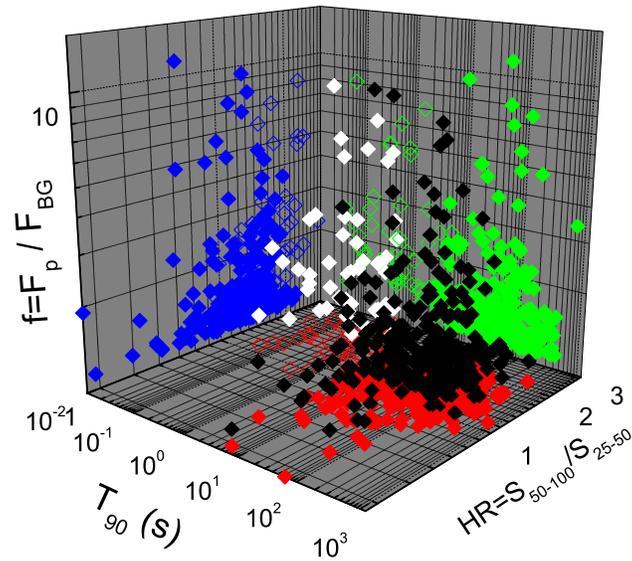}
\caption{The $T_{90}-{\rm HR}-f$ 3D distribution of the {\em
Swift} GRBs in our sample. Long and short GRBs are denoted as
solid black and white diamonds, respectively. Their projections
to the $T_{90}-{\rm HR}$, $T_{90}-f$, and ${\rm HR}-f$ 2D
planes are denoted in red, green and blue colors, respectively,
with the long and short GRBs denoted with solid and open
sympols, respectively.}
\end{figure*}

\begin{figure*}
\includegraphics[angle=0,scale=0.8]{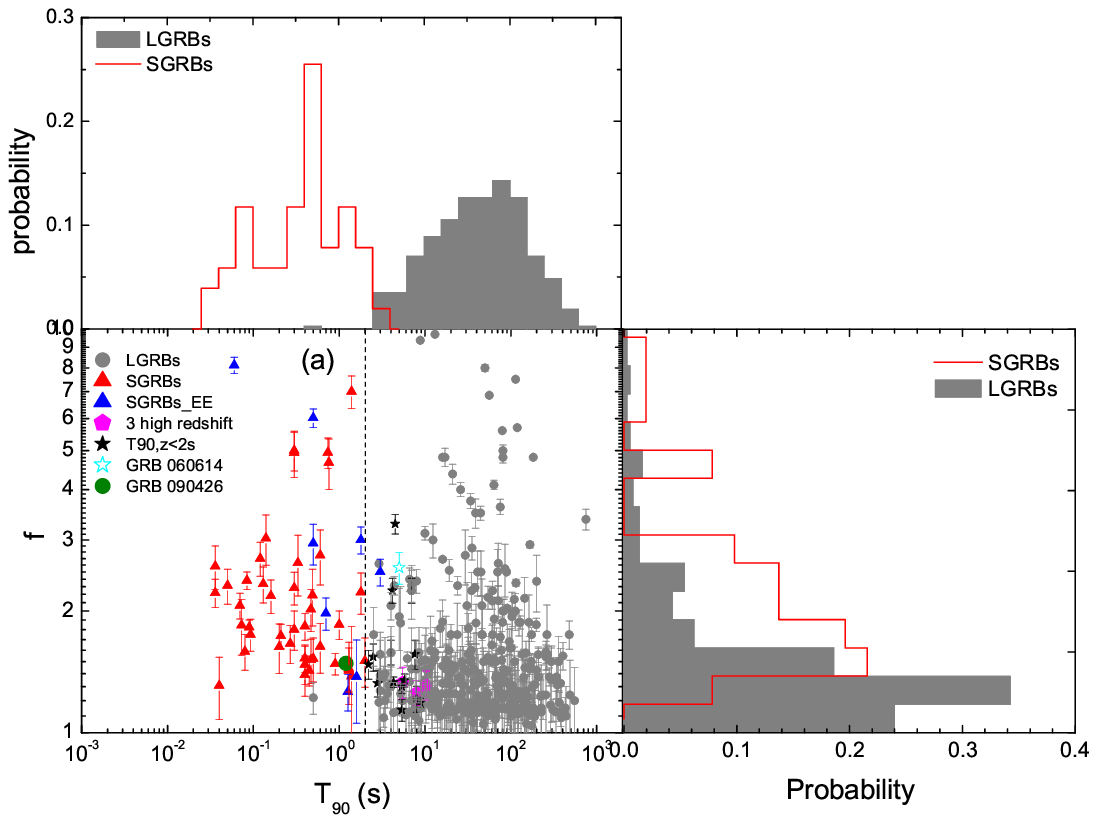}
\includegraphics[angle=0,scale=0.8]{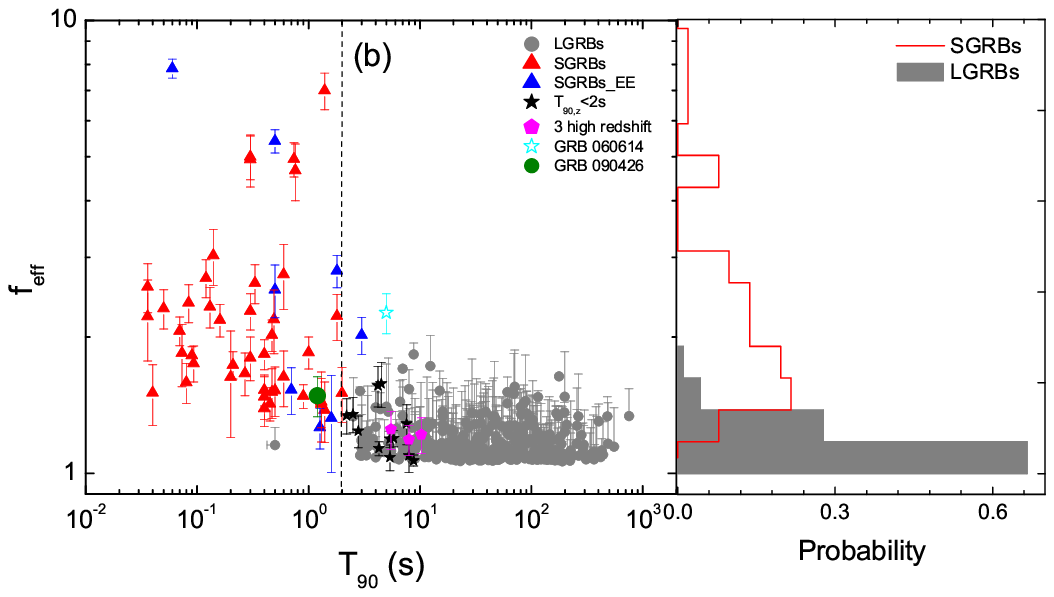}
\includegraphics[angle=0,scale=0.8]{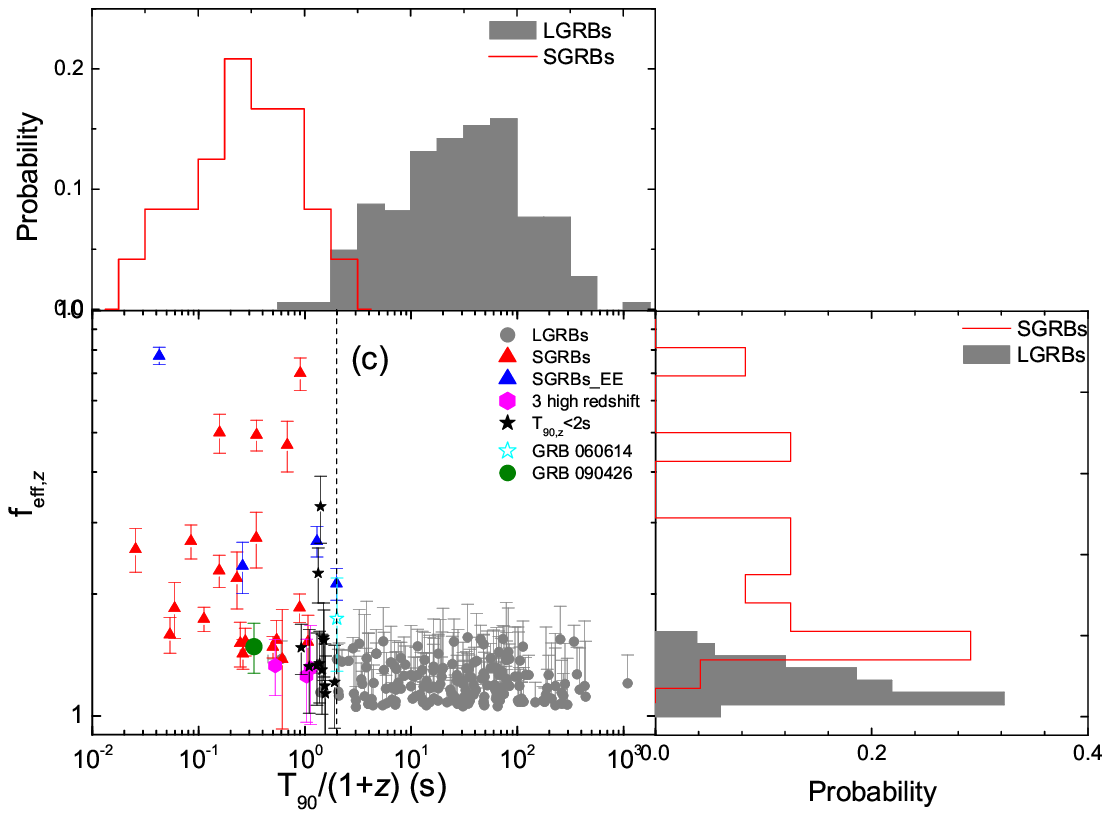}
\caption{The 1D and 2D distributions for the bursts in our
sample. (a): The $T_{90}-f$ diagram of the GRBs in our sample.
(b) The $T_{90}-f_{\rm eff}$ diagram of GRBs in our sample. (c)
The $T_{90}/(1+z) - f_{\rm eff,z}$ diagram of GRBs in our
sample. The following convention is adopted for all three
plots: Gray: long GRBs, red: short GRBs; blue: short GRBs with
extended emission; purple: three GRBs with the highest $z$;
black: ``rest-frame short'' GRBs. GRB 060614 and GRB 090426 are
marked with special symbols. The vertical dashed line is the 2
s separation line.}
\end{figure*}

\begin{figure*}
\includegraphics[width=8cm, angle=0]{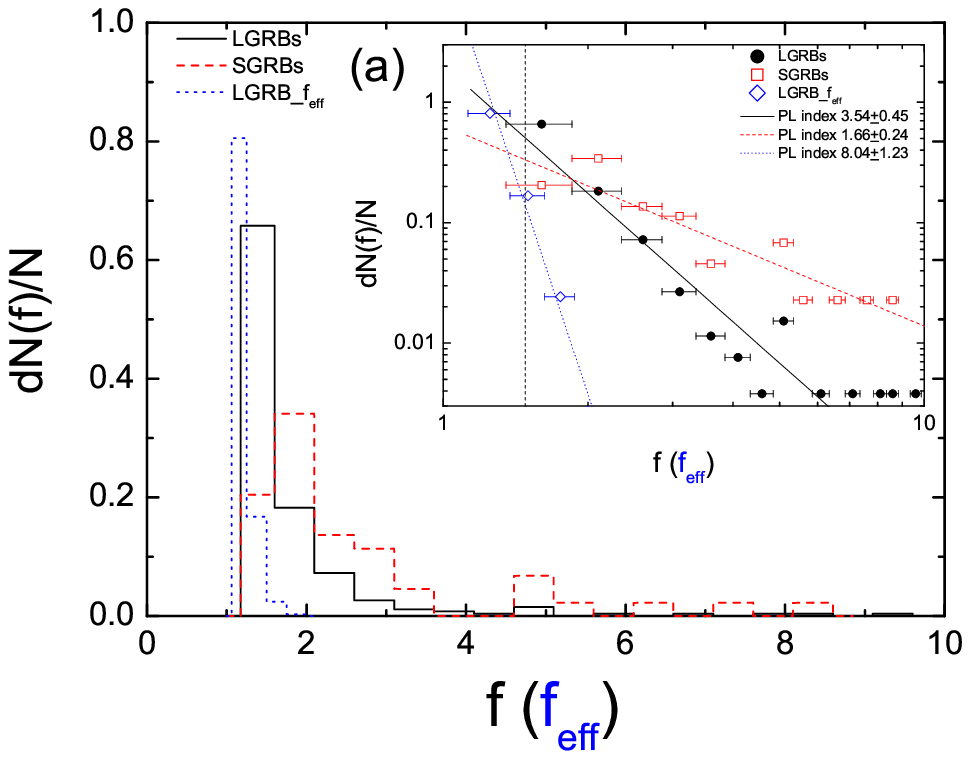}
\includegraphics[width=8cm, angle=0]{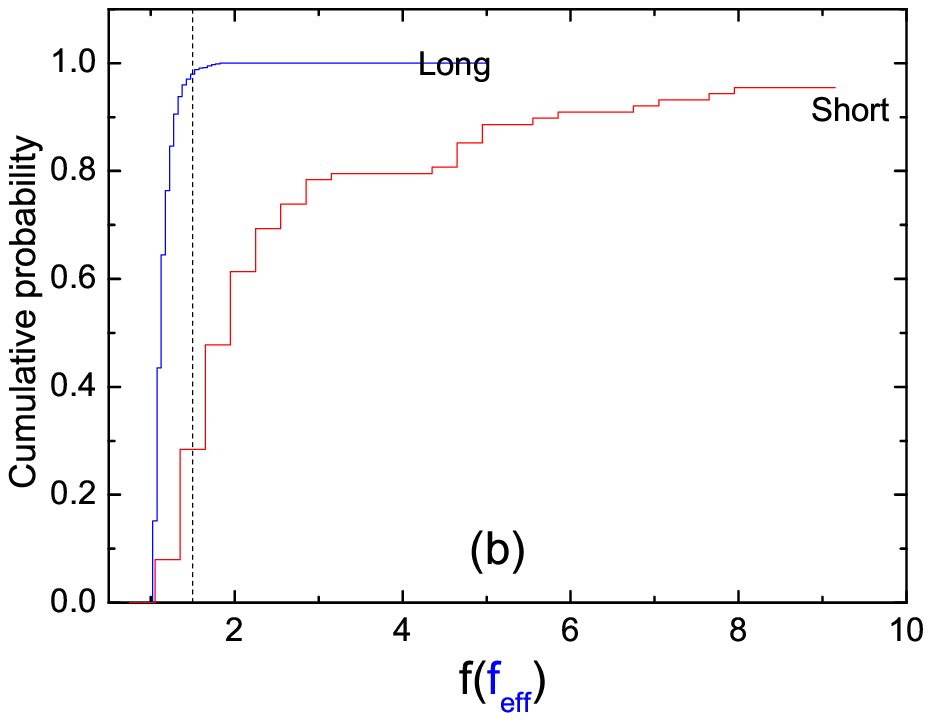}
\includegraphics[width=8cm, angle=0]{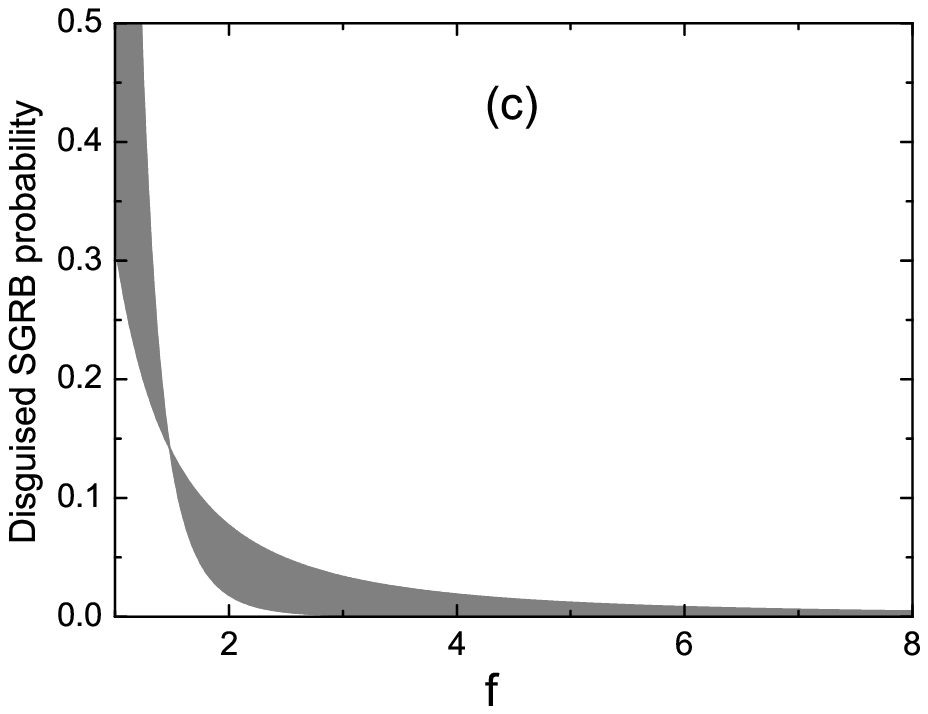}
\caption{(a) The distributions of $f$ (for both long and short
GRBs) and $f_{\rm eff}$ (for long GRBs only) as well as their
power law fits (inset). (b) The cumulative probability of a GRB
below a certain $f$ (for short GRBs) or $f_{\rm eff}$ (for long
GRBs) value. The vertical line corresponds to $f=1.5$. (c)
Chance probability of a disguised short GRB below a certain $f$
value. The gray region is the error zone for the probability.}
\end{figure*}

\begin{figure*}
\includegraphics[width=8cm, angle=0]{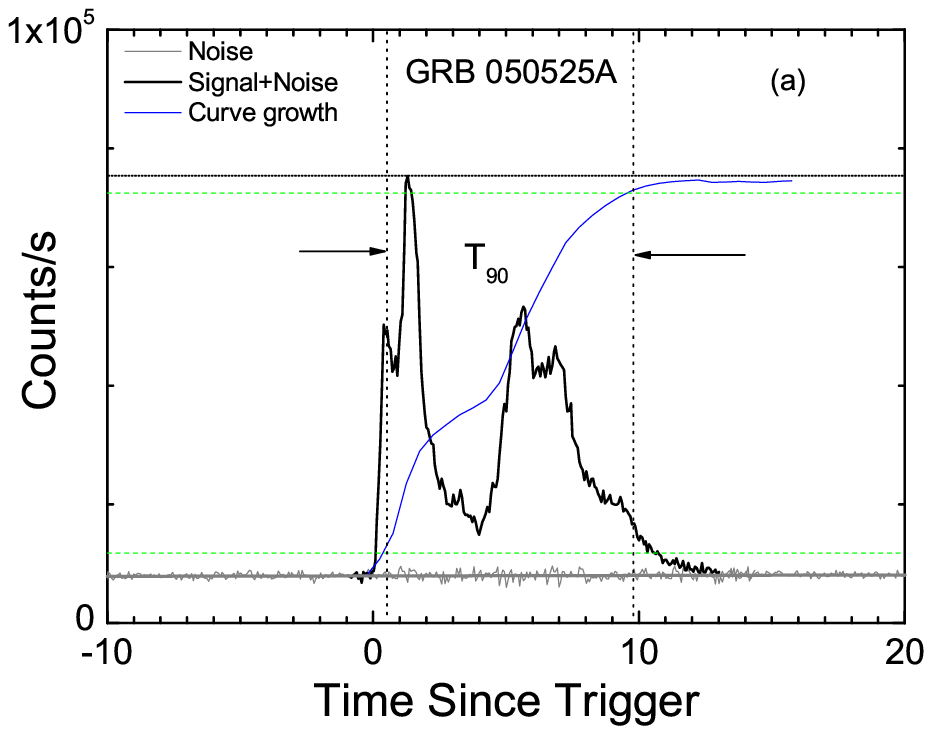}
\includegraphics[width=8cm, angle=0]{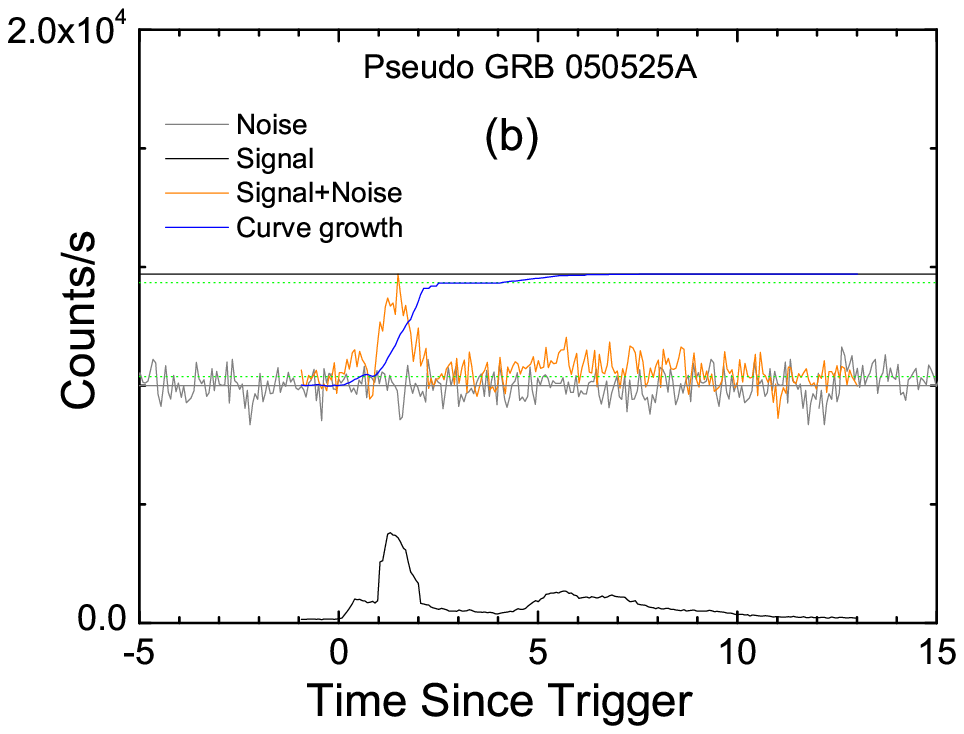}
\caption{\bf An example of defining $f_{\rm eff}$ with GRB
050525A. (a) The original lightcurve and the definition of
$T_{90}$ using the standard ``curve of growth'' method. (b) The
pseudo GRB generated from GRB 050525A. The original lightcurve
is scaled-down by a factor of 0.06 (thin black curve). Adding
the background (grey), the total lightcurve (orange curve) is
the ``observed'' lightcurve of the pseudo GRB. Applying the
curve of growth method, the $T_{90}$ of the pseudo GRB is just
shorter than 2 s. The $f$ parameter of the pseudo GRB, which is
$f_{\rm eff}$ of GRB 050525A, is measure as 1.53.}
\end{figure*}

\begin{figure*}
\includegraphics[width=8cm, angle=0]{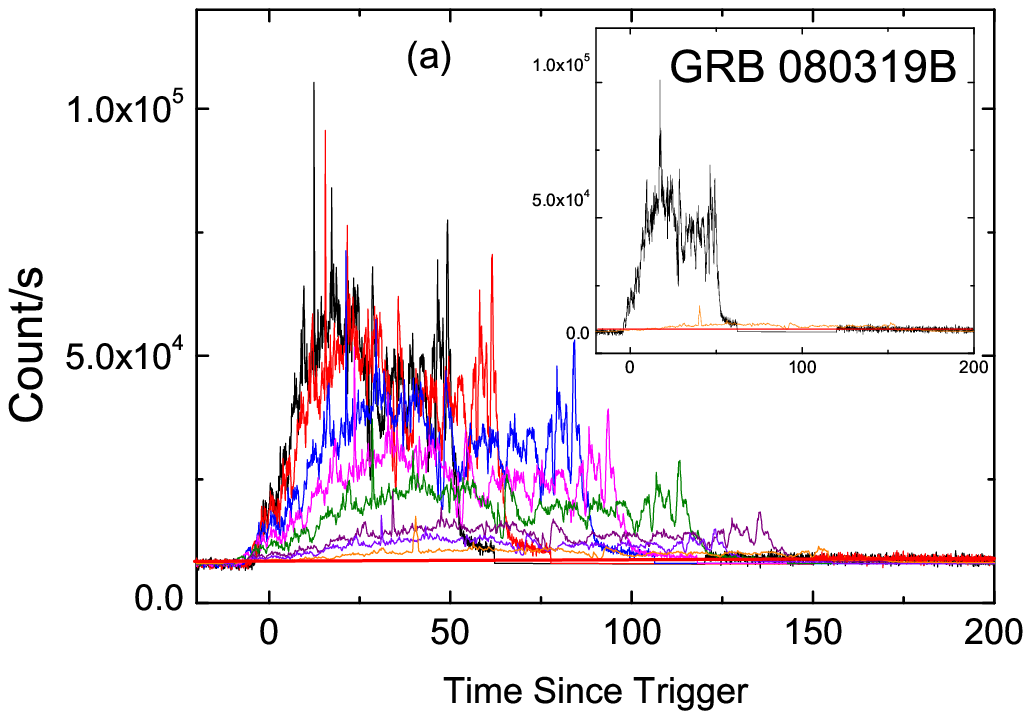}
\includegraphics[width=8cm, angle=0]{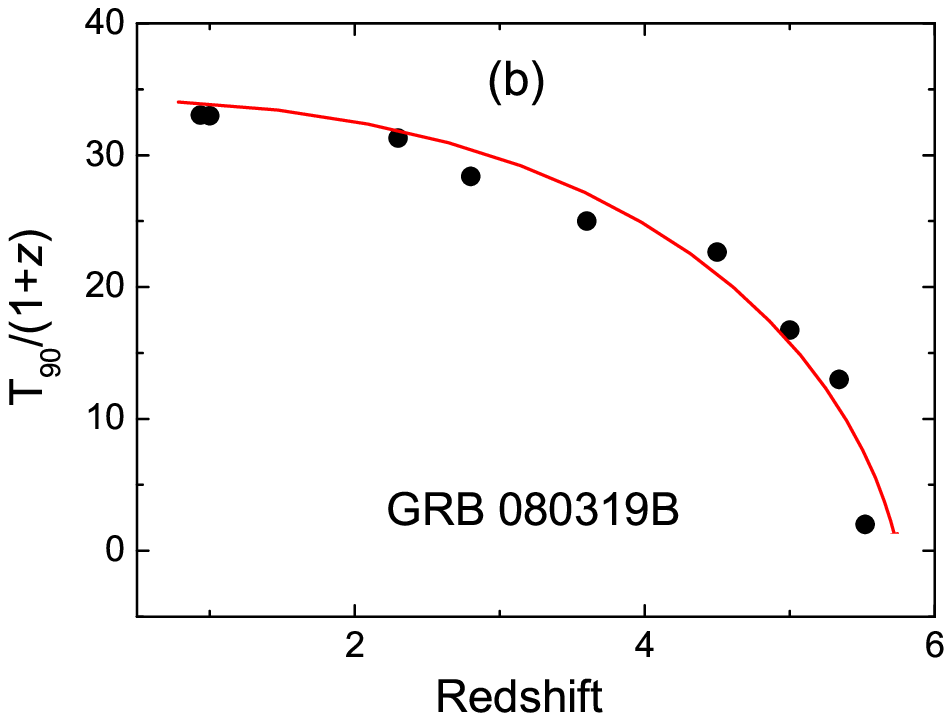}
\caption{(a) The simulated BAT-band pseudo GRB lightcurves by
moving GRB 080319B to progressively high redshifts. From top to
bottom, $z=0.937$, 1, 2.3, 2.8, 3.6, 4.5, 5.1, 5.53. (b) The
measured rest-frame duration $T_{90}/(1+z)$ of the pseudo GRBs
in our simulation. The red solid line shows a smooth broken
power-law fit.}
\end{figure*}

\begin{figure*}
\includegraphics[width=12cm, angle=0]{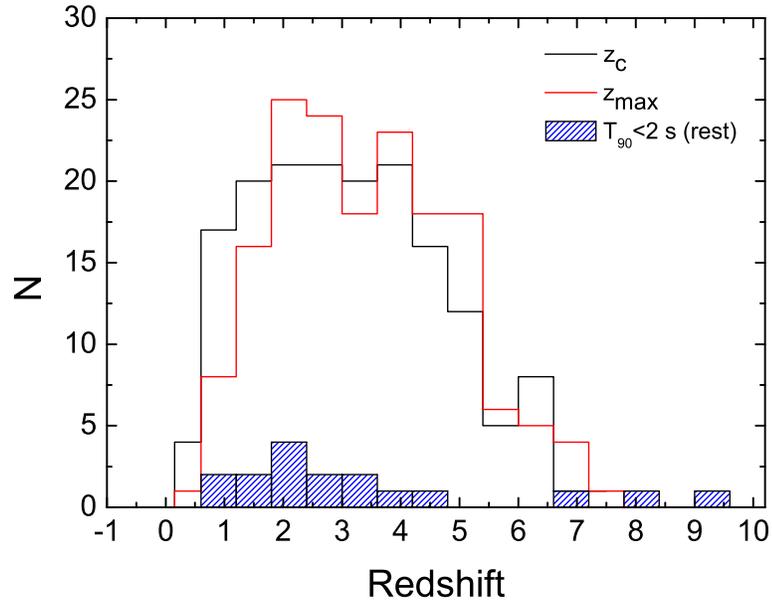}
\caption{The distributions of $z_{c}$ and $z_{\rm max}$ for the
simulated pseudo GRBs as compared with the redshift
distribution of the observed rest-frame short GRBs.}
\end{figure*}

\begin{figure*}
\includegraphics[width=12cm, angle=0]{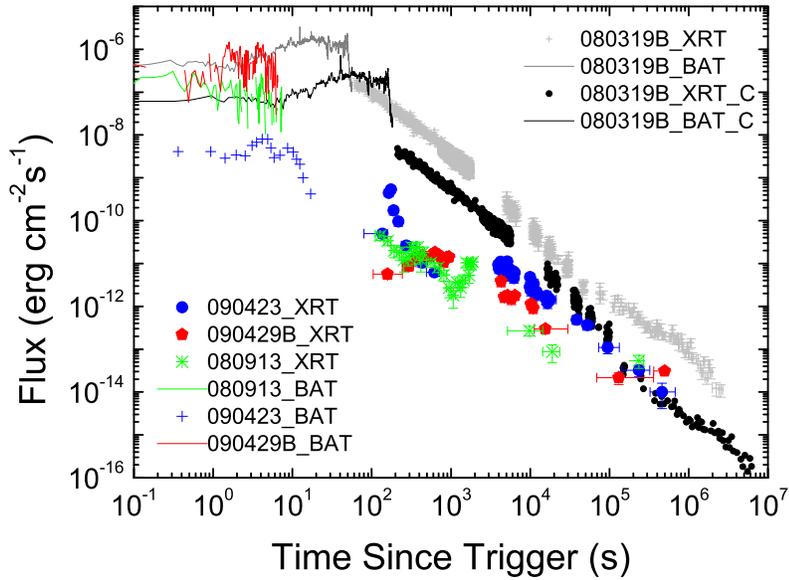}
\caption{The simulated XRT-band lightcurve of the pseudo GRB by
moving GRB 080319B to $z=z_c=5.53$ (black), as compared with
the original XRT-band lightcurves of GRB 080319B (gray), GRB
080913 (green), GRB 090423 (blue), and GRB 090429B (red).}
\end{figure*}


\end{document}